\documentclass{aa}
\usepackage{graphicx}
\usepackage{natbib}
\usepackage{amsmath}
\usepackage{amsfonts}
\usepackage{amssymb}
\usepackage{array}
\usepackage{txfonts}
\usepackage{multirow}
\begin{document}

\title{A Radio Monitoring Survey of Ultra-Luminous X-Ray Sources}

\author{Elmar K\"ording \inst{1},Edward Colbert\inst{2} \and Heino Falcke\inst{1,3,4}}

\institute{Max-Planck-Institut f\"ur Radioastronomie,
           Auf dem H\"ugel 69, 53121 Bonn, Germany\\
           email: koerding@mpifr-bonn.mpg.de
         \and
	John Hopkins University, Department of Physics and Astronomy,
	Homewood Campus, 3400 North Charles Street, Baltimore, MD~~21218
		\and
           Radio Observatory, ASTRON, Dwingeloo, P.O. Box 2, 7990 AA 
Dwingeloo, The Netherlands
         \and
           Dept. of Astronomy, Radboud Universiteit Nijmegen, Postbus 9010, 
 6500 GL Nijmegen,  The Netherlands
}

\date{Oct 2004}

\titlerunning{A Radio Survey of ULXs}

\abstract{ 
We present the results of a radio monitoring campaign to
search for radio emission from nearby ultra-luminous X-ray sources (ULXs). These sources are bright off-nuclear X-ray point sources with luminosities exceeding $L_X > 10^{39}$ erg/sec.
A well-defined sample of the 9 nearest ULXs has been monitored eight times during 5 months with the Very Large Array in A and B configuration. Our limiting sensitivity is $\approx$ 0.15 mJy (4 $\sigma$) for radio flares and $\approx$ 60 $\mu$Jy for continuous emission. In M82 two ULXs seem to have coincident compact radio sources, which are probably  supernova remnants. 
No continuous or flaring radio emission has been detected from any other ULX.
Thus, ULXs do not generally emit steady-state radio emission above radio powers of $1.5 \times 10^{17}$ W/Hz. The non-detections of the continuous emission are consistent with beamed or unbeamed radio emission from accreting black holes of $\leq 10^3 M_\odot$ based on the radio/X-ray correlation.  Other published radio detections (M82, NGC~5408) are also discussed in this context. Both detections are significantly above our detection limit. 
If ULXs have flaring radio emission above $4 \times 10^{17}$ W/Hz we can give an upper limit on the duty circle of the flares of 6\%. This upper limit is in agreement with the observed number of flares in Galactic radio transients. 
Additionally we present a yet unreported radio double structure in the nearby low-luminosity AGN NGC~4736.
}

\maketitle 

\section{Introduction}

Ultra-luminous X-ray sources are among the most widely discussed objects, but their true nature is yet unknown. These objects have X-ray luminosities $L_{\rm X}$ $\sim$ 10$^{39-42}$ erg/s which seems too bright for normal (stellar-mass) black hole X-ray binaries (BHXRBs), but far dimmer than normal active galactic nuclei (AGN). The first hints to these intermediate-luminosity X-ray point sources have been found in the 1980s \citep{Fabbiano1989,ColbertPetreSchlegel1995}.
Observations with ROSAT, and subsequent high resolution X-ray satellites, confirmed these findings and showed that these intriguing sources are often not in the centers of the galaxies \citep{ColbertMushotzky1999,RobertsWarwick2000}. More recent surveys (see e.g., \citealt{ColbertPtak2002}, \citealt{PtakColbert2004}) find approximately one ULX in every five galaxies, confirming that ULXs are a common phenomenon.  

One reason why ULXs are so intensely studied is their potential connection with intermediate mass black holes (IMBHs). Under the assumption that they emit isotropically, the Eddington limit gives a lower bound for the mass of the central black hole as large as a few hundred solar masses for nominal Eddington ratios of 
$\sim$0.1$-$1.
The existence of IMBHs would be extremely exciting, as they could be the 'missing link' between stellar mass black holes and the supermassive black holes in the center of the galaxies (see e.g., \citealt{EbisuzakiMakinoTsuru2001}). 
Such objects would have important implications for cosmology and arise in many theories on the collapse of primordial stars. 

While there is evidence that a few of these objects are indeed 
IMBHs (e.g., \citealt{StrohmayerMushotzky2003}, 
\citealt{CropperSoriaMushotzky2004}), 
the creation and feeding mechanisms for black holes of this mass are 
still not well understood. The ULX phenomenon seems to be connected to star formation, see e.g., the cartwheel galaxy \citep{GaoWangAppleton2003}. This galaxy would need hundreds to thousands of IMBHs feeding from a yet unknown non-stellar mass reservoir or the ULX phenomenon is connected high mass XRBs \citep{King2004}. \citet{GrimmGilfanovSunyaev2002} and \citet{Gilfanov2004} show the existence of a universal X-ray luminosity function that may extent up to ULX luminosities by taking the star forming rate into account.  There seems to be a dependence of the ULX abundance on the galaxy type, since dwarf galaxies host many of the nearby ULXs. 

There is also an ongoing discussion whether the inner accretion disk temperatures of ULXs are too hot for IMBHs, since many have kT$_{in} \sim$ 1~keV, which is
more similar to those found in normal XRBs see e.g., \citet{ColbertMushotzky1999,MizunoOhnishiKubota1999,MakishimaKubotaMizuno2000}. However, recent XMM observations find lower inner disk temperatures for some ULXs, which is more compatible with IMBHs than earlier results from ASCA observations (see e.g. \citealt{MillerFabbianoMiller2003,MillerFabianMiller2004}). 

Besides the explanation that ULXs are accreting IMBHs other stellar-mass ULX models have been proposed:
anisotropic X-ray emission (``mild beaming'') \citep{KingDaviesWard2001}, 
super-Eddington accretion flows \citep{AbramowiczCzernyLasota1988,King2004},
and microblazars \citep{KoerdingFalckeMarkoff2002}. In the microblazar model, the 'ultra-luminous' X-ray emission is created by a relativistically boosted X-ray jet. While the jet is intrinsically weaker than the accretion disk, Doppler boosting can lead to the observed luminosities. If the jets also emit in the radio regime like XRBs in the hard state (cf., \citealt{Fender2001}) or the very high state (see e.g., GRS~1915 \citealt{RodriguezGerardMirabel1995}), the radio emission is also boosted and may be observable. The jet/disk model \citep{FalckeBiermann1995} assumes that the accretion flow and the jet form a symbiotic system; if there is a disk, there is always a jet in some form. Flat spectrum radio emission originates from the self-absorbed part of the jet, while one observes a steep synchrotron power law in the X-rays \citep{MarkoffFalckeFender2001}.  Thus, one possibility to distinguish between the microblazar model and the other possibilities is the presence of compact, luminous radio cores at the positions of ULXs. 

Galactic XRBs in the very high state are probably the nearest cousins to ULXs. Of these GRS~1915+105 can be seen as the prototype. 
GRS~1915+105 shows bright radio flares of 1.5 Jy while the quiescent value is approximately 130 mJy \citep{RodriguezGerardMirabel1995}. An other highly variable XRB is Cyg~X-3. While this source is also found in the low and high state, it shows major radio flares with a flux increase of a factor 10-100 on a time scale of a day which lasts for days or weeks \citep{OgleySpencer2001}.    
Therefore, it has to be expected that also ULXs could be radio transients. These relativistically boosted radio flares could be detectable with the VLA.  In addition, these observations can be used to estimate the number of radio flares created in a galaxy. Such estimates are important for the design of new digital radio telescopes such as LOFAR or the SKA, as they could easily search for transients due to their multi-beam capabilities.

In this paper we describe a systematic search for steady state or flaring radio emission from a well defined sample of ULXs.  
In Sect. 2 we present our observation scheme, and in Sect. 3 we show our 
results.
Theoretical implications are discussed in Sect. 4. Our conclusions are given in Sect. 5. Finally we discuss the low luminosity AGN NGC~4736 in the appendix.

\section{Observations}

\subsection{Sample Selection}

ULXs are defined as having observed 0.3$-$8.0 keV luminosities 
L$_X$ $>$ 10$^{39}$ erg~s$^{-1}$.
We selected nine ULXs from a larger list of $\sim$200 ULXs, which were 
found in Chandra ACIS data that had become public by 4~Feb~2003.
X-ray luminosities were estimated from ACIS count rates using a power-law 
model with photon index $\Gamma =$1.7, and the Galactic absorption column
to the corresponding galaxy (calculated using the FTOOLS NH program).
See Table \ref{tablePositions} for L$_X$ values for the nine ULXs in our sample.

X-ray point source lists, used to create the larger ULX catalog, were 
created using the XASSIST automatic X-ray analysis program
(\citealt{PtakGriffiths2003}, URL: www.xassist.org).
Our nine sample objects are the nearest nine ULXs 
(D $<$ 5.5 Mpc) in this list, with DEC $>$ $-$20$^\circ$, excluding M81 X-1.
The latter source was excluded because it has already been studied with the VLA (more than 4 hours of integration time, no published detections). 
Nevertheless, they represent a {\it distance-limited sample that is unbiased in X-ray luminosity.}

The distance limit was mainly used to reduce the sample size. A low distance also helps to give low limits on the total emitted radio power of the sources. 
This selection results in nine ULXs in seven fields of views (FOVs):
M33, NGC~2403, M82 (2 ULXs), NGC~4736 (2 bright point sources, one may be the nucleus), NGC~5204, and two fields in NGC~5457 which is also known as M101.
The positions of the ULXs are shown in table~\ref{tablePositions}. 
All host galaxies are spirals except M82, which is classified as dwarf/irregular (see e.g., \citealt{deVaucouleursdeVaucouleursCorwin1991}).

\begin{table}
\setlength{\extrarowheight}{1pt}
\caption[]{Positions of the ULXs and their 0.3-8 keV X-ray luminosities. In the fields with only one ULX, the pointing center is the ULX. The field of NGC~4736 was centered at the middle of both sources. A * in front of the ULX name denotes the pointing center. }
\label{tablePositions}
\begin{tabular}{clllc}
\hline \hline
FOV&Name & RA & Dec & X-ray Lum.\\
& & & & $\left[ 10^{39} \frac{\mbox{erg}}{\mbox{s}} \right] $ \\
\hline
1 *&M33 &01:33:50.9 & +30:39:38 & 1 \\     
2 *&NGC~2403 & 07:36:25.5 & +65:35:40  & $2$ \\
\multirow{2}{*}3 *&M82-D & 09:55:50.2 & +69:40:46  & $3$ \\ 
&M82-F & 09:55:51.0 & +69:40:45  &$6$ \\
\multirow{2}{*}{4 *}&NGC~4736-1 & 12:50:53.1 & +41:07:13  &$1$ \\
&NGC~4736-2 & 12:50:53.3 & +41:07:14  & $1$\\ 
5 *& NGC~5204 & 13:29:38.6 & +58:25:06 & $2$ \\ 
6 *& NGC~5457-1 & 14:03:32.4 & +54:21:03 & $2$ \\
7 *& NGC~5457-2 & 14:04:14.3 & +54:26:04 & $1$ \\
\hline
\end{tabular}
\end{table}

\subsection{Observing scheme}
The goal of these observations is to search for radio flares and continuous emission from ULXs using the VLA. The seven selected fields were observed eight times each. The observing runs were approximately four hours long and occurred between June and October 2003. This results in 34 minutes per source and epoch including the calibration scans. After the last epoch the data of all epochs were combined to give deep observations to search for continuous emission. To avoid confusion the observations were obtained in the VLA's highest resolution configurations (A or B configuration). 

The optimal receivers at the VLA to search for weak, flat spectrum point sources are the X-band receivers (8.49 GHz). They are more sensitive than the C and U band receivers, as long as the spectrum is flatter than a spectral index $\alpha$ $(S_\nu \propto \nu^{-\alpha})$ of 0.4. We expect that the ULXs have flat or even inverted spectrum which could be detectable if they are beamed. Thus, we observed at 8.49 GHz using the maximal bandwidth of 50 MHz. Due to the distribution of the right ascension of our sources we had to observe them in two groups (for the groups and the observing dates see table~\ref{tableDates}). 

As we are searching for very faint radio emission, we obtained phase-referenced observations. For all sources VLA phase-calibrators could be found within $10^\circ$, typically within $5^\circ$ of the target source.
The cycle time between calibrator and source was approximately 7 minutes. 
The phase correction between subsequent calibrator scans was typically $\lesssim 40^\circ$ on the longest baselines, while the phase corrections on the shorter baselines are typically $\lesssim 10^\circ$. Therefore, the phase calibrated images should only be degraded by a few percent due to phase errors. Amplitude calibration has been done using either 3C286 or 3C48. 
The phase-calibrated data has then been imaged with natural weighting to achieve the maximal sensitivity. In each epoch a source was observed for approximately 21 minutes, excluding phase and amplitude calibrator observations.

For M82, we have also reanalyzed archival A-configuration data presented in \citet{KronbergSramek1985}, as they have reported a radio flare of 7.07 mJy at C-band (6 cm). Intriguingly, the position of the flare is near the brightest ULX in this galaxy.  

\begin{table}
\caption[]{Observing Dates of the different Epochs. The sources are observed in two groups. The first group consists of M33, NGC~2403, and M82, while NGC~4736, NGC~5204, and the two FOVs in NGC~5457 belong to the second group.}
\label{tableDates}
\begin{center}
\begin{tabular}{cllc}
\hline \hline
Epoch & Group 1 & Group 2 & Configuration\\
\hline
1 & 2 June &  4 Jun & A \\     
2& 3 July &  3 July & A \\
3 & 28 July & 5 July & A\\
4 & 18 Aug & 17 Aug & A\\ 
5 & 7 Sep & 30 Aug & A or AnB \\
\hline
6 & 15. Sep &  15. Sep & AnB\\
7 & 03 Oct & 07 Oct & AnB\\ 
\hline
8 & 16 Oct & 18 Oct & B \\ 
\hline
\end{tabular}
\end{center}
\end{table}

\section{Observational Results}
\begin{table}
\caption[]{RMS flux values for the different FOVs. Besides the RMS of each epoch we give
the RMS value for map of the combined dataset. All values are given in $\mu$Jy. For a discussion on M82 see Sect.~\protect\ref{subM82}}
\label{tablerms}
\begin{tabular}{c|ccccc|ccc|c}
\hline \hline
Epoch & 1 & 2 & 3 & 4 & 5 & 6 & 7 & 8 & comb.\\
\hline
M33 & 37 & 36 & 37 & 37 & 60 & 41 & 71 & 54 & 17 \\
2403 & 31 & 30 & 32 & 31 & 59 & 38 & 45 & 37 & 13\\
N5457I1 & 33 & 32 & 31 & 35 & 35 & 39 & 77 & 46 & 14\\
N5457I2 & 32 & 32 & 34 & 37 & 33 & 35 & 75 & 34 & 13\\
N5204 & 35 & 33 & 31 & 35 & 34 & 37 & 51 & 38 & 13\\
N4736 & 45 & 47 & 49 & 55 & 45 & 61 & 63 & 62 & 21\\
\hline
\end{tabular}
\end{table}

\subsection{Error limits}
The root mean square (rms) flux values of the observed sources have been calculated from natural weighted maps. For fields with detected sources the largest possible rectangular field excluding the source has been used. The theoretical rms value for a naturally weighted 21 minute observation with 27 antennae is, according to the VLA Observational Status Summary \citep{TaylorUlvestadPerley2004}, 32 $\mu$Jy.
The rms flux values of the individual epochs and fields are shown in table~\ref{tablerms}. In a single epoch the typical observed rms is about 35 $\mu$Jy (except for M82, see Sect.~\ref{subM82}). If one combines all epochs, the rms goes down to $\approx 15 \mu$Jy, while the theoretical limit is 11 $\mu$Jy. 
To be able to interpret the radio images we have to ask when a peak in the map is significant. Chandra has a point spread function (PSF) of approximately one arc-second.
\citet{WrobelTaylorGregory2001} report that astrometric error for phase referenced VLA observations (A-configuration) compared to the International Celestial Reference Frame as established by VLBI \citep{MaAriasEubanks1998} is less than 10 mas for the X-band. 
The astrometric uncertainty of the VLA is therefore negligible compared to the Chandra position errors.

We know the positions of the ULXs with the accuracy of Chandra. During the first five epochs the VLA was in A configuration and our beam width was $\approx 240$ mas. The PSF of Chandra which has a full width half maximum of around one arcsecond, is therefore covered by
17 beams. The probability that we detect a random peak in 17 independent beams is given by the error function. Under the assumption that these beams are indeed independent, the chance that there is a random 3 $\sigma$ peak at the position of a given ULX is 4 $\%$, the chance of a 4 $\sigma$ peak 0.1 $\%$. We can therefore only accept peaks at the positions of the ULXs, if they exceed 4 $\sigma$. To search for other radio flares in the whole map we have to increase the 4 $\sigma$ limit.
We typically map 50" by 50", which corresponds to roughly 14000 'independent' beams, so there will usually be one 4 $\sigma$ peak in the map. Therefore point sources without known positions are not believable unless they have a higher signal-to-noise than $\approx 5 \sigma$.

\subsection{Detections}
\subsubsection{M82}\label{subM82}
\begin{figure*}
\resizebox{17cm}{!}{\includegraphics[angle=0]{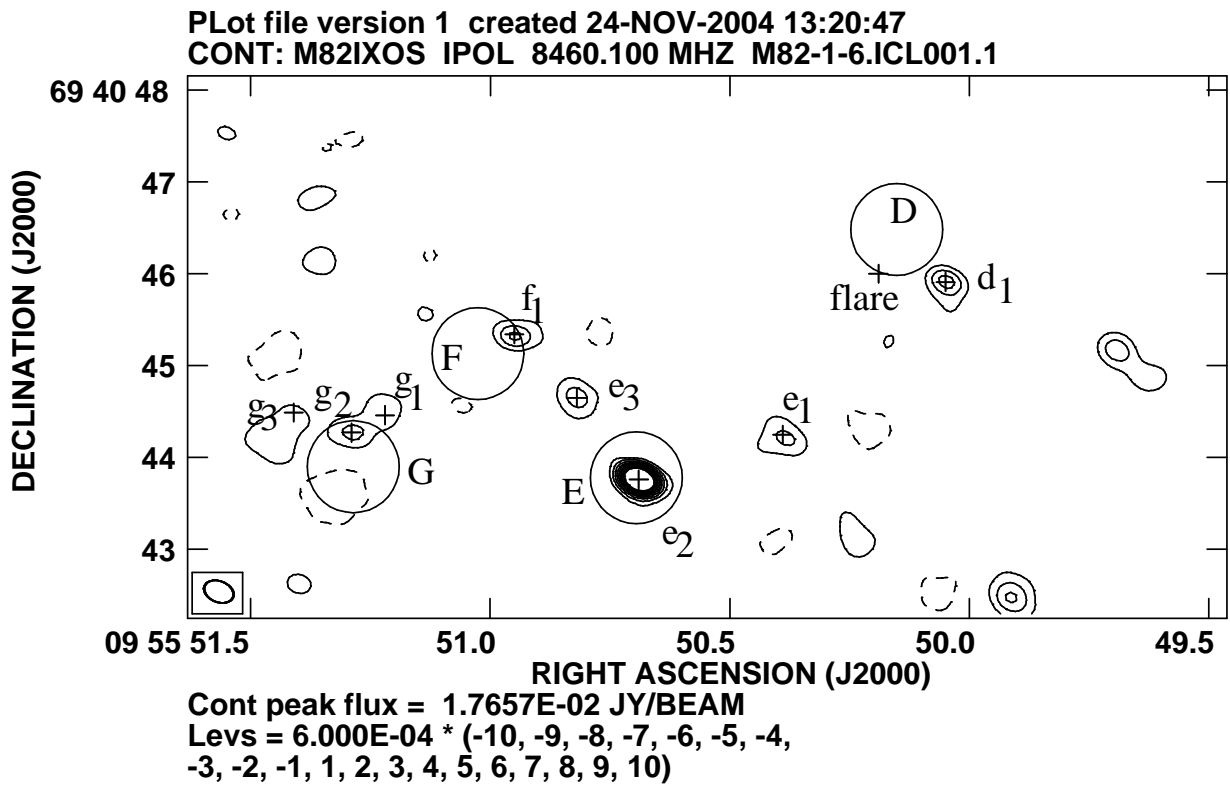}}
\caption{Map of M82 around the ULXs D and F (see table~\ref{taM82det}). The brightest radio source in the field is $e_2$ which is coincident with X-ray point source E. The FWHM of all X-ray point sources have been set to 1". Even though the radio source $d_1$ is outside of the r=0.5" circle for ULX D, its offset (0.77") is consistent with the ULX X-ray position.} 
\label{fiM82}
\end{figure*}
In the starbursting galaxy M82 many compact radio sources (CRSs) have been found (see eg., \citealt{KronbergWilkinson1975}). Many of them have been identified as supernova remnants (SNRs). This has been done by direct imaging (e.g., \citealt{PedlarMuxlowGarrett1999}) or by spectral observations (see \citealt{AllenKronberg1998}). M82 also shows strong diffuse radio emission, making it more difficult to achieve the optimal noise limit for the short A-config snap-shot observations (zero-spacing, dynamic range limitations). The longer baselines are less contaminated by this diffuse emission. Thus, the best rms values have been obtained by only using baselines exceeding 300 k$\lambda$ for the A and early AnB config observations. This omits approximately half of the baselines, increasing the theoretical rms by a factor of $\approx \sqrt{2}$ to 44 $\mu$Jy.  The rms values for a single epoch still do not reach this value but are a bit higher: $\approx 70$ $\mu$Jy. For the 6th to the 8th epoch smaller cut-offs had to be utilized while the array configuration approached B-configuration. In table \ref{taM82det} we give the X-ray and radio positions. While the X-ray positions are named with capital letters, the corresponding radio detections are named as indexed small letters. The X-ray sources D and F are the well known ULXs \citep{MatsumotoTsuruKoyama2001}.

Our VLA observation was centered on ULX D of table \ref{taM82det}. Thus, the bright X-ray point source A is 18.8" away from the phase tracking center. As we are observing in continuum mode, our sensitivity is already reduced for this ULX due to bandwidth smearing. The distance corresponds to 72 beam widths, thus, the peak flux will be reduced by 5$\%$ \citep{TaylorUlvestadPerley2004}. We have not detected any radio emission from source A.

The brightest CRS in the FOV ($e_2$, 41.95+57.5 in the notation of \citealt{MuxlowPedlarWilkinson1994}, abbreviated to MPW) is in good agreement with the position of a bright X-ray point source (E). This source is probably a SNR (e.g., \citealt{AllenKronberg1998}) that may be immersed in a molecular cloud \citep{PedlarMuxlowGarrett1999}.  As the radio and X-ray positions coincide, the radio and X-ray maps are correctly aligned.  

Near the ULX D we have detected a CRS ($d_1$) with a continuum flux of 1.8 mJy in all eight epochs observed in this campaign. This source is also detected by MPW with 1.8 mJy as 41.3+59.6 and is identified as a SNR  \citep{AllenKronberg1998}. As expected, the flux of the source is stable during the 8 epochs within $1 \sigma$ of the rms. It has a distance of 0.77" to the ULX, well in agreement with the Chandra position. We did not detect the flaring source of \citet{KronbergSramek1985} in our VLA observations.

For the ULX F we find two possible radio counterparts as shown in table \ref{taM82det}. The brighter one of both is also given in earlier papers as 42.21+59.0 while the weaker one is barely visible in the maps of MPW but not given in their table. Both sources are not variable over our observing period.

Besides the CRS near the reported positions of the ULXs we have not found any radio flare in M82. However, \citet{KronbergSramek1985} found a 7.07 mJy flare near the brightest ULX (D) in M82 ($2.3 \times 10^{19}$ W/Hz). We have reanalyzed their data for direct comparison with our results. The position of the M82-flare is  is 0.56" away from the Chandra position of the ULX, see table \ref{taM82det}. It is already visible in the phase-referenced images before any self-calibration. In our re-analyzed maps the flux of the flaring source is slightly higher ($\approx 9$ mJy) than the value found by \citet{KronbergSramek1985}.
The brightest CRS at that time ($e_2$) had a flux of 105 mJy in 1981 (108 mJy as measured by \citealt{KronbergSramek1985}), currently this CRS has an X-band flux of 12.3 mJy. 
The M82-flare is stable throughout the 7 hour observation in 1981, and is also visible if one divides the observation into 1 hour blocks.  
Even though, the detected radio flare is near the ULX position, it can well be, that this source is also a SNR. The radio decay timescale and the spectrum is similar to the SN 1983n in M83 \citep{KronbergSramekBirk2000}, which was identified with a SN associated with a massive star \citep{SramekPanagiaWeiler1984}. 

\begin{table}
\setlength{\extrarowheight}{1pt}
\caption[]{Radio detections near detected X-ray point sources in M82. A flux of 1 mJy corresponds to $3 \times 10^{18}$ W/Hz at the distance of M82.
}
\label{taM82det}
\begin{tabular}{lllll}\hline \hline
Name & X-ray Position & Radio F. & Dist.\\
&\ \ \ Radio Position & [mJy] & arcsec\\
\hline
A & 09:55:46.59 +69:40:40.9 \\
B & 09:55:46.71 +69:40:37.9 \\
C & 09:55:47.48 +69:40:59.6 \\
D & 09:55:50.15 +69:40:46.5 \\
\ \ \ $d_1$ & \ \ \ 09:55:50.05 +69:40:45.9 & 1.8 & 0.77\\
\ \ \ M82-flare & \ \ \ 09:55:50.19 +69:40:46.0 & 9.0 & 0.56\\
E & 09:55:50.70 +69:40:43.8 \\
\ \ \ $e_1$& \ \ \ 09:55:50.39 +69:40:44.3 & 1.5 & 1.7\\
\ \ \ $e_2$& \ \ \ 09:55:50.69 +69:40:43.8 & 12.3 & 0.06\\
\ \ \ $e_3$& \ \ \ 09:55:50.82 +69:40:44.7 & 1.6 &1.08\\
F & 09:55:51.03 +69:40:45.1\\
\ \ \ $e_3$& \ \ \ 09:55:50.82 +69:40:44.7 & 1.6 & 1.21\\
\ \ \ $f_1$ & \ \ \ 09:55:50.95 +69:40:45.3 & 1.9 &0.47\\
G & 09:55:51.29 +69:40:43.9 \\
\ \ \ $g_1$& \ \ \ 09:55:51.22 +69:40:44.5 & 1.3 & 0.67 \\
\ \ \ $g_2$& \ \ \ 09:55:51.29 +69:40:44.3 & 1.3 & 0.37 \\
\ \ \ $g_3$& \ \ \ 09:55:51.41 +69:40:44.5 & 0.4 & 0.87 \\
H & 09:55:51.48 +69:40:36.0 \\
I & 09:55:53.43 +69:41:02.0 \\
\hline
\end{tabular}
\end{table}

\begin{table}
\setlength{\extrarowheight}{1pt}
\caption[]{Other radio detections}
\label{taRaddet}
\begin{tabular}{lllll}\hline \hline
Name & Radio Position & Radio Flux [mJy] \\
\hline
\multicolumn{3}{c}{NGC~4736}\\
N4736-$a$ & 12 50 53.03  +41 07 13.6 &1.6\\
N4736-$b$ & 12 50 53.08  +41 07 13.0 &1.2\\
\hline
\multicolumn{3}{c}{M101} \\
M101-$a$ & 14 03 31.31  +54 21 14.9 & 0.092\\
M101-$b$ & 14 04 14.32  +54 26 09.4 &0.057 \\
M101-$c$ & 14 04 14.25  +54 26 09.4 &0.060 \\
\hline
\hline
\multicolumn{3}{c}{Low significance flux peaks, see discussion} \\
\hline
M101-$\alpha$ & 14 04 14.21  +54 26 02.6 & 0.15  \\ 
N2403-$\alpha$ & 07 36 25.49 +65 35 39.8 & 0.14 \\
\hline
\end{tabular}
\end{table}

\subsubsection{NGC~4736} \label{ssngc4736}
\begin{figure}
\resizebox{8cm}{!}{\includegraphics{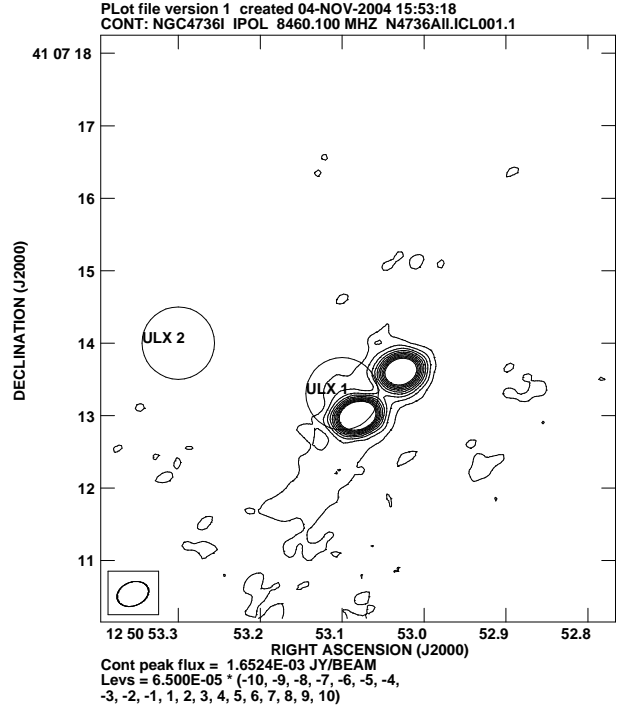}}
\caption{The double source in NGC~4736. The upper source is called source a and the lower and stronger source is called b.}
\label{fiN4736Map}
\end{figure}

During all epochs we detect a double radio source near ULX NGC~4736-1 as shown in Fig.~\ref{fiN4736Map}. Their separation is about 1", which corresponds to 20 pc at the galaxy for a distance of 4.3 Mpc. The position of the stronger radio source coincides with the position of the LLAGN reported by \citet{NagarFalckeWilson2002} and is in agreement with the Chandra position of ULX NGC~4736-1 (see table \ref{tablePositions} and Fig.~\ref{fiN4736Map}). No radio emission is detected at the position of the other ULX NGC~4736-2 (see Fig.~\ref{fiN4736Map}). As one of the two X-ray sources is probably the nucleus, we assume that the stronger radio source of the two (source $b$) is the nucleus, while the other radio source is of unknown nature. As this double structure is probably not related with the ULX phenomenon we discuss the findings in the appendix A. 

\subsubsection{NGC~5457 (M101)} \label{subM101}
In the combined maps of both fields of view the galaxy NGC~5457, which is also known as M101, in we find emission at a 5  $\sigma$ level (M101-$a$), which is not at the positions of the ULXs.
At the same position is a $4 \sigma$ peak in the eighth epoch. If one excludes this epoch from the map the source is still visible but has a reduced flux of 72.9 $\mu$Jy. 
In the second FOV in NGC~5457 there seems to be a double source: M101-$b$ and M101-$c$. Both sources are $5 \sigma$ detections. The nature of these sources is unknown.

\subsection{Non-detections}
There are no significant detections found in M33, NGC~2403, and NGC~5204 in the individual epochs or the full data with a signal to noise higher than 4.
In M101 we do not find significant detection in the individual epochs, for continuous emission see Sect.~\ref{subM101}. Here we limited the search radius to 2" from the ULX Chandra X-ray positions.

In NGC~2403 we found in the fourth epoch a peak in the map (N2403-$\alpha$) which is a 4.4 sigma detection. The VLA position coincides with the ULX position within the accuracy of Chandra. 
However, the emission seems to originate only from a short time period (5 of 21 minutes). If these times are omitted, the peak in the map vanishes. Some XRBs show such short time radio variability \citep{OgleySpencer2001}. No second short radio flare was seen in our data. So, it is not possible to confirm such a short flare, especially as the significance is only marginally. We therefore do not consider this 'flux peak' as a detection of a flare.

In M101 we detect a 4.2 $\sigma$ peak (M101-$\alpha$, see table \ref{taRaddet}) in the second VLA pointing centered on the ULX NGC~5457-2. The radio source is 1.4" away from the Chandra position of the ULX, further away than the typical position accuracy of Chandra, but it is still possible that it is connected to the ULX. However, besides this peak there is another similar maximum another arcsecond away, and both maxima are found on a sidelobe. Thus, it is possible that the source could be an instrumental artifact. Besides this, we do not detect significant ($> 4 \sigma$) radio emission from the ULXs in the individual epochs.

\begin{table}
\setlength{\extrarowheight}{1pt}
\caption[]{The radio power of our sample for continuous emission and flares. The radio detections in M82 are probably SNRs while in NGC~4736 we probably detect the nucleus of the galaxy. For all other sources we present upper limits for continuous emission over all epochs and flares. The distances are taken from \protect\citet{Tully1988}.}
\label{tableRadPower}
\begin{tabular}{llllc}
\hline \hline
Name & Dist. & L$_{\mbox{X}}$ & L$_{\mbox{R,flare}}$ & L$_{\mbox{R,cont}}$\\
 & [Mpc] & $\left[ 10^{39} \frac{\mbox{erg}}{\mbox{s}} \right] $ & $\left[ 10^{33} \frac{\mbox{erg}}{\mbox{s}} \right] $ & $\left[ 10^{33} \frac{\mbox{erg}}{\mbox{s}} \right] $ \\
\hline
M33 & 0.7 & 1 & $< 0.43$  & $< 0.2$ \\     
NGC~2403 & 4.2 & $2$ & $<13$ & $<5.4$ \\
M82-D & 5.2 & $3$ & $< 45$ & 290 \\
M82-F & 5.2 &$6$ & $< 45$ & 307 \\
NGC~4736-1 &4.3 &$1$ & $< 21$ & 177 \\
NGC~4736-2 & 4.3 & $1$ & $< 21$ & $< 9.2$  \\ 
NGC~5204 & 4.8& $2$ & $< 18$  & $< 7.1$ \\ 
NGC~5457-1 &5.4& $2$ & $< 23$ & $< 9.1$ \\
NGC~5457-2 &5.4& $1$ & $< 23$ & $< 9.1$ \\
\hline
\end{tabular}
\end{table}

\section{Theoretical Interpretations}

\subsection{Flares}
The nearest known Galactic cousins of ULXs are the highly accreting black hole XRBs. XRBs in the low hard or the high state are highly variable in the radio and X-ray regime (c.f., \citealt{Klis1989}). The radio flux can vary by a few hundred  percent, for example Cyg X-1 has a typical 15 GHz radio flux of approximately 10 mJy and reaches up to 30 mJy \citep{Pottschmidt2002}. As already mentioned in the introduction stronger accreting objects like GRS~1915+105 or Cyg X-3 show violent radio flares of a factor 10 to 100. During these flares 
GRS~1915+105 reaches 1.5 Jy \citep{RodriguezGerardMirabel1995} and Cyg~X-3 goes up to more than 15 Jy \citep{WatanabeKitamotoMiyamoto1994}, while the 'steady' state emission is around 100 mJy for both sources. The brighter radio bursts of  Cyg~X-3 may be due to a higher Doppler factor, as the inclination of this source is only $14^{\circ}$ \citep{MioduszewskiRupenHjellming2001}. GRS~1915+105 on the other hand is seen nearly edge on (angle to the line of sight: $66^{\circ}$,  \citealt{RodriguezGerardMirabel1995}). 

One possibility to estimate the luminosities of radio flares one could expect from beamed emission of ULXs is by taking the transients GRS~1915+105 and Cyg X-3 as an example. The average distance of our ULXs is approximately 4.6 Mpc. GRS~1915+105 has a distance of 11 kpc \citep{FenderGarringtonMcKay1999} while Cyg X-3 is 9 kpc away \citep{PredehlBurwitzPaerels2000}. Without any relativistic beaming these sources would have a flux of 8.5 $\mu$Jy and 57 $\mu$Jy, which is below our detection limit. However, already a mild beaming factor of 20 for GRS~1915 or an additional factor of 3 for Cyg X-3 would bring the flares into our detection limit (0.15 mJy). As GRS~1915 is seen with an inclination angle of $66^{\circ}$, its Doppler factor will be around one. Thus, a beaming factor of 20 can be reached with a moderate Lorentz factor of $\Gamma = 3$ \citep{LindBlandford1985}, if the jet points roughly at the observer (inclination angle $ < 15^{\circ}$). The Doppler factor of Cyg X-3 is uncertain, as the observed luminosity is probably already beamed. But an additional factor of 3 is easy to obtain if the jet is pointing directly to the observer. If the ULXs are flaring in radio we should be able to detect the beamed radio flares.

We have not found a single significant flare for all our sources in the monitoring campaign. In the individual epochs our $4 \sigma$ sensitivity is on the average 0.15 mJy.

The distance of M33 is around 0.84 Mpc \citep{FreedmanWilsonMadore1991}. Besides this nearby galaxy all other observed galaxies have distances in the range from 3.6 Mpc to 5.4 Mpc. Thus, the detection limits of flares will only vary by a factor of two for those galaxies. For the average distance of 4.6 Mpc the upper limit on the radio power $S_\nu$ of flares is $3.8 \times 10^{17}$ W/Hz. This corresponds to a 5 GHz radio luminosity $(\nu S_{\nu})$ of $1.9 \times 10^{34}$ erg/sec assuming a flat spectrum. For M33, however, the limit is reduced to a radio power of $1.1 \times 10^{16}$ W/Hz and a radio luminosity of $5.7 \times 10^{32}$ erg/sec.

The non-detection may be due to the unknown time scales of the radio flares in ULXs. We observed our sources once or twice a month. If the time scale of a radio flare is only a day, we are strongly under-sampling the radio light-curve.
The time scale of the boosted flares is unclear and depends on the physical process creating the flare. If the flare is created inside the jet, e.g., similar to the shock in jet models used for blazars (see e.g., \citealt{MarscherGear1985}) , the observed time scale of the flare will be reduced by the Lorentz factor. For $\Gamma \approx 5$ the time scale could be as short as a few hours. However, it will be extremely bright, even a few mJys are possible.  On the other hand, if the flare is created by enhanced injection of material into the jet by the disk for an extended time, the observed time scale will be the same as the intrinsic one. 

The fact that we have not found a single flare in all our epochs can be translated to an upper limit for the probability that an average ULXs of our sample has a radio flare brighter than $3.8 \times 10^{17}$ W/Hz, which should be detectable. 
If we assume that the probability that an ULXs is flaring at a given time is similar for all observed ULXs, and one flare is uncorrelated to earlier or later flares, the flares should be Poisson distributed. 
The probability distribution describing how many events we detect given the probability of a detection is
\[P(\lambda,n) = \frac{\exp^{-\lambda} \lambda^n}{n!}, \]
where $\lambda$ denotes the expectation value of the distribution and n is the number of events.

For this study we have to exclude M82 as the rms in the M82 maps is much higher than in the other fields. The other six fields have been observed for eight times which yields 48 samples. Let $\rho$ denote the upper bound of the duty cycle of the ULX. Thus, it is an upper limit for the probability that one ULX flares at a given moment.  $\rho$ has to be chosen, such that we should have almost certainly detected at least one event if the probability that a given ULX flares is $\rho$. Here $\rho$ will be chosen such that we should detect one or more flares with a probability of 95$\%$. This leads to 
\[ 0.95 = \sum_{n=1}^{n=\infty} P(48 \rho,n) = 1-\exp^{-48 \rho} . \]
This results in $\rho = 0.06$, i.e. the duty cycle of radio flares exceeding our detection threshold in ULXs is $< 6 \%$. The expectation value for the number of flares is 2.9, our detection of no flares is in agreement with this value at a 5 $\%$ level.  

If we know the time scale of a typical radio flare in an ULX we can convert the upper limit of the flaring probability to an upper limit of the number of flares a ULX can have in a year. 
Let $\delta t_{\mathrm{Flare}}$ denote an average length of a flare. As the radio flares of GRS~1915 have a time scale of days, we will use  $\delta t_{\mathrm{Flare}} \approx 2$ days as a reference. Thus, the upper limit of the probability to detect a flare in a single observation corresponds to an upper limit of 11 $\frac{\delta t_{\mathrm{Flare}}}{2 \mathrm{days}}$ radio flares per year. This upper limit is still higher than the number of bright flares in GRS~1915+105.

Besides looking for radio emission from ULXs we can use these observations to derive upper limits on the amount of radio transients happening in a galaxy. 
Our galaxies have an average distance around 4.6 Mpc and we mapped an area of 51", this results in an observed area of 1.3 kpc$^2$.
We are searching for radio flares from unknown sources, therefore, it is also not known which is the quantity correlating with the number of flares. While for ULXs the star formation rate might be a good quantity it could simply be the number of stars or black holes for an other class of flaring objects. As we are observing a couple of small fields in different galaxies, it is very hard to derive the exact number of observed solar masses or even the observed star formation rate.
As most of our observed galaxies (besides M82, which we exclude) are non starbursting galaxies, the star formation rate per solar mass will be of the same order of magnitude.
In order to get a rough estimate how much mass we have actually observed, we assume a similar mass density as in our Galactic neighborhood.
In our Galactic neighborhood the mass density of the disk is of the order of $200\  \rm{M}_{\odot}$ per pc$^2$. Therefore one field of view observes a mass of approximately $3\times 10^8\ \rm{M}_{\odot}$. Thus, we expect less than 0.2 flares in a single observation of $10^9\  \rm{M}_{\odot}$.

\subsection{Background sources}
We have detected continuous radio sources only inside NGC~5457 (excluding M82). In all other fields there were no background sources visible. However, we only map an area of $51"\times51"$. 
\citet{FomalontKellermannPartridge2002} find that the number density of sources above a flux density $S$ is approximately
\begin{equation}
N = 0.099\pm 0.01 \left(\frac{S}{40 \mu\mbox{Jy}} \right)^{-1.11\pm 0.13} \mbox{arcmin}^{-2}.
\end{equation}
Our limiting sensitivity $(4 \sigma)$ is approximately 56 $\mu$Jy.
Thus, this predicts that there should on average be $0.049\pm0.007$ sources in a given observed $51"\times51"$ map. Our findings are in agreement with this result. The probability that a background source is found coincident with a ULX, within the position accuracy of Chandra (1 arcsec), is below $2\times10^{-5}$.   

\subsection{Steady State Emission}
\begin{figure*}
\resizebox{17cm}{!}{\includegraphics{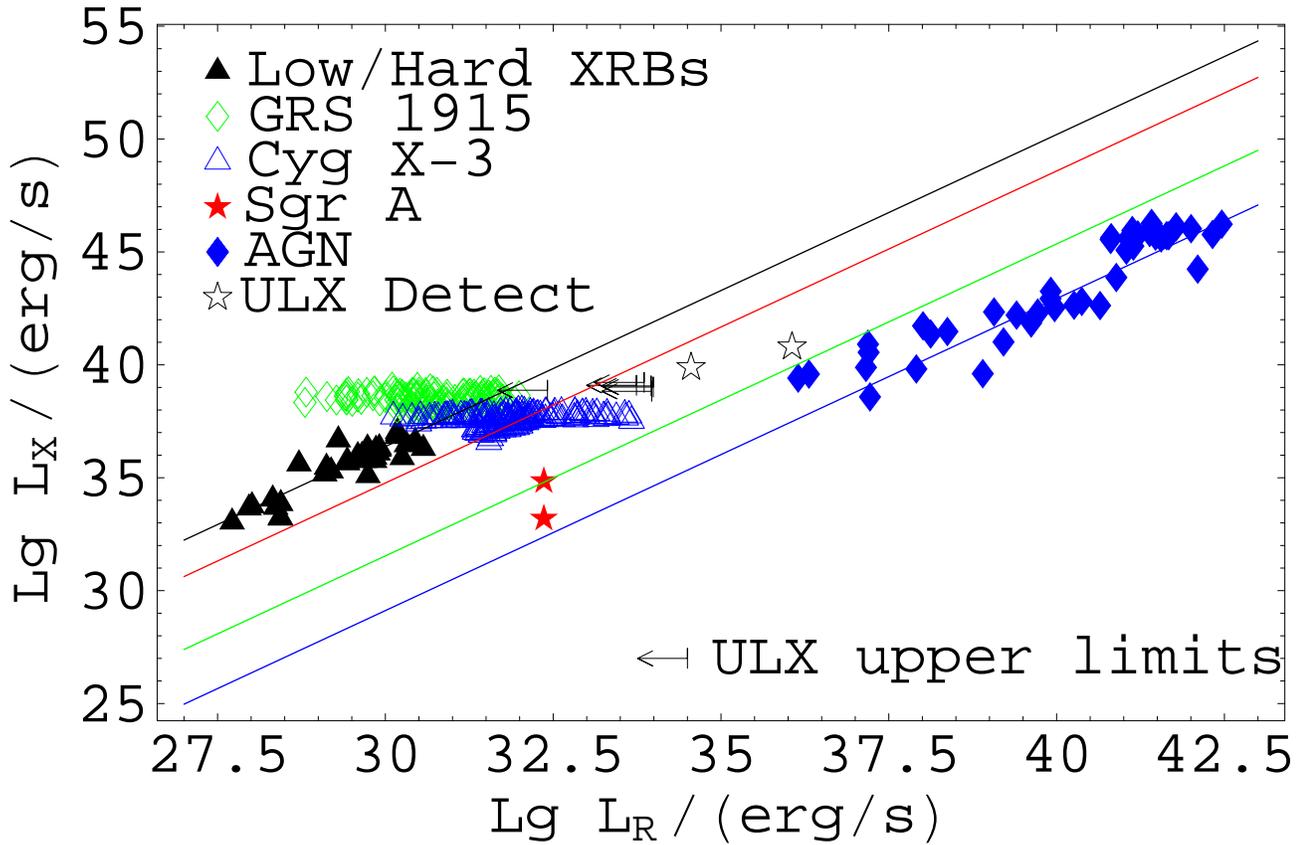}}
\caption{The upper limits of the ULXs in respect to the low/hard state XRB and AGN population. The lines correspond to different black hole masses. From Top to bottom $6, 6\times 10^2, 6\times 10^6, 6 \times 10^9{\rm M}_{\odot}$. The left ULX radio detection represents NGC~5408, while the right is the radio flare in M82. The correlation should not hold for fares, this was just plotted for comparison. As GRS~1915 and Cyg X-3 are both radio transients, they are not expected to follow the correlation.}
\label{fiULXCorr}
\end{figure*}

In M82 two reported Chandra ULXs positions are in agreement with the positions of CRSs, that are probably radio SNR. For all other galaxies we have not detected any continuous emission near the ULXs. The upper flux limit for the observed sources is $\approx 60 \mu$Jy ($4 \sigma$). 
For the steady state emission ULXs should lie on the Radio/X-ray correlation, which was first  established for XRBs \citep{CorbelFenderTzioumis2000,CorbelNowakFender2003,GalloFenderPooley2003} and has been extended to AGN \citep{MerloniHeinzdiMatteo2003,FalckeKoerdingMarkoff2004}.
Simple jet scaling (see e.g., \citealt{FalckeKoerdingMarkoff2004}) predicts that the X-ray and Radio emission are correlated and scale according to: 
\begin{equation}
L_{\rm X} \propto L_{\rm R}^{1.38} M^{0.81}
\label{eqRXScale}
\end{equation}
where $L_{\rm X}$ and $L_{\rm R}$ denote the X-ray and radio luminosities. 
The exact value of the exponents depends on the spectral indices of the Radio and X-ray emission. In Fig.~\ref{fiULXCorr} we plot the 3--9 keV X-ray luminosity against the 5 GHz radio luminosity. The radio luminosity is always calculated assuming a flat spectrum, i.e., it is defined as $L_{\mbox{5 GHz}} = 5 GHz S_{\mbox{5 GHz}}$ where $S_{\mbox{5 GHz}}$ denotes the radio power at 5 GHz. The Chandra X-ray luminosities are corrected for the different energy band assuming a photon index of 1.6. Besides a sample of low hard state XRB (GX339-4, V404 Cyg, XTE J1118+480, 4U 1543-47, GS 1354-64) we show the two XRBs used to extrapolate to the expected radio flare luminosities: GRS1915 and Cyg X-3. The data-points have been taken from \citet{GalloFenderPooley2003}. The brightest 1.5 Jy flares of GRS1915 are not included in the plot. Additionally we present a sample of AGN consisting of low luminosity AGN, FR-I radio galaxies \citep{FanaroffRiley1974}, and Radio and X-ray selected BL Lac objects. This figure is similar to Fig.~3 in \citet{FalckeKoerdingMarkoff2004}. Furthermore, we show the theoretical radio/X-ray scaling for Black holes of different black hole masses.

A strict radio/X-ray correlation will only be valid for the steady state emission of low/hard state, jet-dominated sources. For high state objects the radio emission will be quenched (see, e.g., \citealt{FenderCorbelTzioumis1999,GalloFenderPooley2003}). For one possible correlation extending to the very high state see \citet{MaccaroneGalloFender2003}.
Furtheremore, we can not expect that the correlation holds for the flares of the transient sources. Those flaring sources are only included for comparison.

The empirical correlation can be used to compare the upper limits ($4 \sigma$) of this radio monitoring campaign with the expectations. The radio flux of high or very high state objects may be reduced compared to the low/hard state objects.  The correlation gives an upper bound for the expected radio flux.
Only the upper limit for M33 is close to the value expected from the correlation for stellar mass black holes. All other upper limit are far above the expected radio flux from microblazars with stellar mass black holes. 
Our non-detections are consistent with either beamed or non-beamed radio emission from stellar or intermediate mass black holes that are $\lesssim 10^3 {\rm M}_{\odot}$.

\subsection{Other radio detections of ULXs}
Up to now there are two radio detections mentioned in the literature, here we discuss their interpretation within the microblazar model.
\citet{KaaretCorbelPrestwich2003} have detected radio emission from an ULX with an X-ray luminosity $1.1\times 10^{40}$ erg/sec in NGC~5408. They find a 4.8-GHz radio flux of 0.26 $\pm$ 0.04 mJy well above our detection limit (17 $\sigma$ for continuous emission). Assuming a distance of 4.8 Mpc \citep{KarachentsevSharinaDolphin2002} this corresponds to a radio power $7.2 \times 10^{17}$ W/Hz, which is also above the upper limits for our sample of ULXs ($1.5 \times 10^{17}$ W/Hz).  
For the X-band they give an upper limit of 0.12 mJy (3 $\sigma$).
The spectral index is therefore larger than $\alpha > 1.0$. We would expect that the radio cores of microblazars have a flat to inverted spectrum, by far flatter than the observed steep spectrum. The radio source does not seem to be a transient as it was detected with a roughly similar flux in a subsequent observation (Kaaret priv. comm.).
As we have the radio and X-ray luminosity we can use the radio/X-ray correlation to compare the ULX in NGC~5408 with XRBs and AGN. The source seems to be too radio loud for a stellar mass black hole, as shown in Fig.~\ref{fiULXCorr}. Under the assumption that the source should lie on the correlation, we can derive a mass for the object of $\sim 1000 M_{\odot}$. Otherwise, the observed emission could be explained by a flare, or by extended emission like a radio supernova. At least this object seems to be different than the ULX sample observed by us.

\citet{KronbergSramek1985} found a 7.07 mJy flare in M82 (L$_R = 1.1 \times 10^{36}$ erg/sec ), which we also confirmed by reanalyzing their data. The flare position is within the errors of the brightest ULX in M82 which has up to $9\times 10^{40}$ erg/sec. This flare is bright compared with GRS~1915, which we expect to show boosted 0.4 mJy flares. However, the value does not seem to be unrealistic for a major flare and the uncertainty of the flux estimate. Up to now no second event has been reported, even though M82 has been frequently observed. Thus, these events seem to be extremely rare. This is consistent with our conclusion that the duty circle of ULX flares in our sample is very low ($<6$\%). For comparison we have also shown the flare in Fig.~\ref{fiULXCorr}. However, the radio/X-ray correlation is only valid for steady state emission.

\section{Conclusions}
We have monitored nine ULXs during five months with the VLA. Our limiting sensitivity (4 $\sigma$) of the individual epochs is $\approx 0.15$ mJy ($ 4 \times 10^{17}$ W/Hz for average distance, rms $\approx 36 \mu$Jy), with the exception of M82. We have not found any significant radio flare in the sample. If they last only a few days, we are heavily under-sampling the light-curve. However, we can give an upper bound of 6\% for the probability to find a flare brighter than $4 \times 10^{17}$ W/Hz in a given observation of an ULXs.
This translates to an upper bound of 11 $\frac{\delta t_{\mathrm{flare}}}{2 \mathrm{days}}$ flares per year with a time scale $\delta t_{\mathrm{flare}}$. This is much more than observed for example in GRS~1915. Thus this monitoring campaign can not rule out or strengthen the idea of relativistically beamed flares in ULXs.

We have reconfirmed the 7 mJy flare near the brightest ULX in M82 in archival data from \citet{KronbergSramek1985}. This corresponds to a radio power of $2.3 \times 10^{19}$ W/Hz. Such a radio flare can be well understood in the context of relativistically beamed emission from the ULX.
An other possible explanation is a SN similar to SN 1983n \citep{KronbergSramekBirk2000}.

We have detected a yet unreported variable double source in NGC~4736. As the positions of the stronger source coincides with the position of the nucleus of the LLAGN as reported in \citet{NagarFalckeWilson2002}, we assume that this double source is probably connected to the LLAGN and not to the phenomenon of ULXs.

The search for continuous emission from ULXs was only successful for M82 (limiting flux: $\approx$ 60 $\mu$Jy, $4 \sigma$). Here two CRSs (probably SNRs) could be associated with ULXs. Thus, ULXs do not generally emit steady-state radio emission above radio powers of $1.5 \times 10^{17}$ W/Hz. Using the Radio/X-ray correlatio we have shown that our non-detections are consistent with either beamed or non-beamed radio emission from stellar or intermediate mass black holes that are $\lesssim 10^3 {\rm M}_{\odot}$.
The search for a radio loud ULXs is therefore still open. Our results suggest that detections will be extremely rare for current telescopes. Nevertheless, the investigations of these intriguing sources have to continue. 
 
{\it Acknowledgments: } The National Radio Astronomy Observatory is a facility of the National Science Foundation operated under cooperative agreement by Associated Universities, Inc.  The European VLBI Network is a joint facility of European, Chinese, South African and other radio astronomy institutes funded by their national research councils. E.K. was supported for this research through a stipend from the International Max Planck Research School (IMPRS) for Radio and Infrared Astronomy at the University of Bonn. 
EJMC acknowledges support from NASA grant NNG 04-GA26G. 

\appendix
\section{The LLAGN in NGC~4736}
\begin{figure}
\resizebox{8cm}{!}{\includegraphics{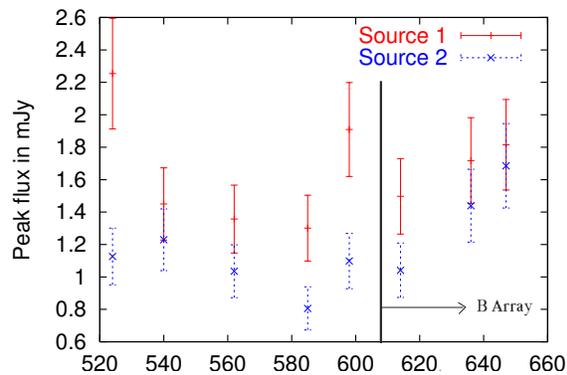}}
\caption{Light-curve of NGC~4736. Note that the VLA array configuration changed from A to B during the last epochs.}
\label{fiN4736Light}
\end{figure}

As we have already mentioned in Sect.~\ref{ssngc4736}, we have detected a double source in NGC~4736 show in Fig.~\ref{fiN4736Map}. One of these sources is stronger (2.2-1.2 mJy, source $b$) 
and variable while the other source seems slightly extended and is fairly stable in flux  (1 mJy, source $a$). The positions are given in table~\ref{taRaddet}
The peak fluxes and the ratio of the two sources is shown in table~\ref{tableNGC4736}, for a light-curve of the peak fluxes see Fig.~\ref{fiN4736Light}.

Due to the low fluxes of both sources it is not possible to self-calibrate the data itself. The fluxes can therefore be reduced by phase errors. They can furthermore be changed by amplitude calibration. However, the flux-ratio of the two sources should be independent of these effects. In the radio the variability can also be seen, it varies between 2 and 1.2 on a time scale of months as also shown in the table.  We therefore conclude, that at least the stronger source $b$ is variable. This indicates that source $b$ is a compact object, probably the nucleus of the low-luminosity AGN. It has the position that was previously published by \citet{NagarFalckeWilson2002}. 
Source $a$ is not mentioned in earlier publications, but \citet{NagarFalckeWilson2002} rms value was significantly higher (0.34 mJy). We have reanalyzed their data, and detect a 4 $\sigma$ peak at the position of source $a$ (1.4 mJy). Source $a$ was therefore already visible in March 1998.  

For the discussion of the other source we have to consider that the array configuration of  the VLA changed between the epochs five and eight from A-configuration to B-configuration, increasing 
the beam size from 260 mas to 840 mas. Extended emission will therefore contribute more to later epochs than to the first. This explains the increasing peak fluxes of the source $a$ from 1 mJy to 1.7 mJy in the last epochs. The effect that the peak flux increases with the beam-size can be seen if one uses a uvtaper of 400 $k\lambda$, which roughly doubles the beam-diameter. The peak flux increases from 1.1 to 1.3 mJy in epoch one.
The extension of source $a$ is also visible in the difference between the peak and the integrated flux. For source $b$ these two values are usually similar, while for source $a$ the integrated value is higher by approximately 30 $\%$. 

It is puzzling that the integrated fluxes are more variable (1.1 -- 1.8 mJy) than the peak fluxes. We checked that this effect is not due to phase errors, epoch one and two have phase jumps between successive calibrator scans of less then $40^{\circ}$, usually less than $10^{\circ}$. This would result in only a few ($<6\%$) amplitude loss. One problem of the integrated flux is that the integration boxes are small, as both point sources are less than an arc-second away. Another possibility could be that source $a$ is barely resolved, small changes in the beam size and the atmosphere could than lead to the observed variability. 
An extended ($>200$ mas) source which is variable on time scales of months at a distance of 4 Mpc is not possible, as its size would be larger than 5 pc. However, there could be a compact substructure in the second source which is varying.

During the last epoch we also obtained a C-band map of NGC~4736. As the VLA was already in B-configuration the two sources could not be separated. Both sources combined give a peak flux of 4.1 mJy and an integrated flux of 7.3 mJy.

To check the spectrum of the two sources we have also obtained U-band images during the last epoch. Source b has a peak and integrated flux flux of 1.7 mJy. The weaker source two has a peak flux of 1.1 mJy and 0.95 mJy integrated flux. The missing integrated flux is probably due to the small integration box around the peak needed to avoid source $b$. The diameter of the beam was 480 mas.
We have reanalyzed the old data from Mar 1998 to check for the second source. Both peaks are visible in the map with fluxes of 2.235 mJy and 1.403 mJy respectively. However, the rms was 0.34 mJy making accurate flux measurements impossible.

We can not compare the U-band flux directly to the X-band flux, as the beam size of the X-band observations is about twice the U-band beam size. As the flux of the second source is roughly constant during the first epochs, we compare the U-band flux with the last epoch in A-configuration and taper the data to get a beam-size of 480 mas. The flux of the second source increases to 1.41 mJy.
This results in a spectral index between 15 and 8.4 GHz of $\alpha = 0.42$.

To check whether these sources are compact we have observed NGC~4736 with the European VLBI Network. While source $b$ was detected with a correlated flux of 0.86 mJy source $a$ was not detected. The rms in this observation was 0.1 mJy.

No other low-luminosity AGN (LLAGN) in 2 cm survey of \citet{NagarFalckeWilson2002} shows such a double structure. The nature of the second source is unknown. 
\citet{MaozFilippenkoHo1995} report for this galaxy a second UV source 2.5" away from the nucleus (see also \citealt{MaozNagarFalckeWilson2005}) and suggest that this galaxy could be in the final stage of a merger. We have not detected their second source. While the position angle of the UV source is -3$^{\circ}$ the second radio source is at $-49^{\circ}$.
We conclude that source $a$ could either be a hotspot or a jet feature of the LLAGN. The spectral index is fairly flat. Thus, the hotspot has to be self-absorbed. If this source would be observed at a greater distance and the two components unresolved, the spectrum would peak at a few GHz, as source $b$ has a flat spectrum. This source could therefore be a close-up of a GHz peak source (GPS).
An other explanation is that inside the extended emission there is a compact flat spectrum source below the detection limit (0.3 mJy) of our EVN observations. If confirmed, this would be spectacular as this could either be a second nucleus or the radio core of an ultra-luminous X-ray source.

\begin{table*}
\setlength{\extrarowheight}{1pt}
\caption[]{Radio flux and beam size of the double source in NGC~4736}
\label{tableNGC4736}
\begin{tabular}{ccccccc}
\hline \hline
Julian Date & Peak 1 & Integ. 1 & Peak 2 & Integ. 2 & Ratio Peak & Beam size \\
& [mJy] & [mJy] &[mJy] &[mJy] & & [mas] \\
\hline
59524 & 2.25 & 2.1 & 1.13 & 1.4 & 2.0 & 251 \\
59540 & 1.45 & 1.7 & 1.23 & 1.8 & 1.2 & 279 \\
59562 & 1.36 & 1.4 & 1.04 & 1.3 & 1.3 & 261 \\ 
59585 & 1.30 & 1.3 & 0.81 & 1.1 & 1.6 & 268 \\
59598 & 1.91 & 2.0 & 1.10 & 1.7 & 1.7 & 260 \\
59614 & 1.50 & 1.6 & 1.04 & 1.3 & 1.4 & 414 \\ 
59636 & 1.72 & & 1.44 & & 1.2 & 504 \\
59647 & 1.82 & & 1.69 & & 1.1 & 827 \\
\hline
\end{tabular}
\end{table*}

\bibliography{refs} 
\bibliographystyle{aa}

\end{document}